\documentclass[english,aps,preprint,superscriptaddress]{revtex4}
\usepackage[T1]{fontenc}
\usepackage[latin9]{inputenc}
\usepackage{amsmath}
\usepackage{esint}

\makeatletter
\@ifundefined{textcolor}{}
{%
 \definecolor{BLACK}{gray}{0}
 \definecolor{WHITE}{gray}{1}
 \definecolor{RED}{rgb}{1,0,0}
 \definecolor{GREEN}{rgb}{0,1,0}
 \definecolor{BLUE}{rgb}{0,0,1}
 \definecolor{CYAN}{cmyk}{1,0,0,0}
 \definecolor{MAGENTA}{cmyk}{0,1,0,0}
 \definecolor{YELLOW}{cmyk}{0,0,1,0}
}

\makeatletter

\newcommand{\be}{\begin{equation}}\newcommand{\ee}{\end{equation}}\newcommand{\ba}{\begin{align}}\newcommand{\ea}{\end{align}}\def\bea{\begin{eqnarray}}\def\eea{\end{eqnarray}}

\usepackage{slashed}

\makeatother

\usepackage{babel}
\begin{document}

\title{Microcausality of spin-induced noncommutative theories}

\author{Jia-Hui Huang}

\email{huangjh19@163.com}

\selectlanguage{english}%

\affiliation{Center of Mathematical Science, Zhejiang University, Hangzhou, P.R.
China}

\affiliation{Department of Physics, Zhejiang University, Hangzhou, P.R.China}

\affiliation{Kavli Institute for Theoretical Physics China, CAS, Beijing 100190,
China}

\author{Weijian Wang}

\thanks{corresponding author}

\email{wjnwang96@gmail.com}

\selectlanguage{english}%

\affiliation{Department of Physics, Zhejiang University, Hangzhou, P.R.China}

\affiliation{Kavli Institute for Theoretical Physics China, CAS, Beijing 100190,
China}
\begin{abstract}
In this brief report, the microcausility of quantum field theory on
spin-induced noncommutative spacetime is discussed. It is found that
for spacelike seperation the microcausality is not obeyed by the theory
generally. It means that Lorentz covariance can not guarantee microcausality
in quantum field thoery. We also give some comments about quantum
field thoeries on such noncommutative spacetime and the relations
between noncommutative spacetime and causality.
\end{abstract}
\maketitle

\section{Introduction}

One of the most important problems in theoretical physics is the quantization
of gravity. There are several approaches in this direction, such as
string theories and loop quantum gravity. Although we are still far
from a complete theory of quantum gravity, some common notions are
obtained. For example, at small scale or Planck scale, the smooth
Riemann manifold structure of the large spacetime will disappear and
the spacetime will manifest some quantum or discrete properties. The
first discrete spacetime model was suggested by Snyder \cite{snyder1947}
at 1947 and physical models on noncommutative spacetime are studied
extensively during last decades. Here we refer two reviews \cite{Douglas2001,Szabo2003}.
The most common noncommutative spacetime model is the canonical model,
where the spacetime operators satisfying the following relations

\begin{equation}
[\hat{x}^{\mu},\hat{x}^{\nu}]=i\theta^{\mu\nu},\label{eq:canonc}
\end{equation}
where $\theta^{\mu\nu}$ is constant noncommutative parameters. This
model is not only a simple one but can be derived from string theory
when there are some backgroud gauge fields on branes \cite{dou,cheung1998,chu1999,seiberg1999}.
Using Moyal-Weyl correspondence, noncommutative field $\phi(\hat{x})$
defined on eq.(\ref{eq:canonc}) can be described by common field
$\phi(x)$ with Moyal product \cite{moyal}, 
\begin{equation}
\phi_{1}*\phi_{2}(x)=\exp(\frac{i}{2}\theta^{\mu\nu}\partial_{\mu}^{x}\partial_{\nu}^{y})\phi_{1}(x)\phi_{2}(y)|_{y\rightarrow x}.\label{eq:moyalproduct}
\end{equation}
We encode the noncommutativity or nonlocality into the Moyal product.
The above noncommutative models have one problem that it breaks Lorentz
invariance explicitly. In \cite{Chaichian:2004yh,Wess:2003da}, the
authors suggested a twisted Lorentz symmetry to resolve this problem. 

Recently, an interesting Lorentz covariant noncommutative model is
suggested \cite{Falomir:2009cq,Gomes:2010xk}. Some phenomenological
applications of this model are also been studied\cite{Das:2011tj}.In
this model, noncommutativity is related with the spin of the fields.
For a scalar particle, no spacetime nocommutativity can be felt. But
a Dirac particle can feel it. The commutation relations of this model
for a Dirac field are 

\begin{eqnarray}
[\hat{x}^{\mu},\hat{x}^{\nu}] & = & -i\theta\epsilon^{\mu\nu\rho\sigma}\hat{S}_{\rho\sigma}+\frac{i}{2}\theta^{2}\epsilon^{\mu\nu\rho\sigma}\hat{W}_{\rho}\hat{p}_{\sigma},\label{eq:spinnc1}\\
{}[\hat{x}^{\mu},\hat{p}^{\nu}] & = & i\eta^{\mu\nu}.\label{eq:spinnc2}
\end{eqnarray}
$W_{\rho}=\frac{1}{2}\epsilon^{\mu\nu\rho\sigma}\hat{S}_{\rho\sigma}\hat{p}_{\nu}$
is the Pauli-Lubanski vector operator. And $\hat{S}_{\rho\sigma}$
are commutators between Dirac $\gamma$ matrices and defined as
\begin{equation}
\hat{S}_{\rho\sigma}=-\frac{i}{4}(\gamma_{\rho}\gamma_{\sigma}-\gamma_{\sigma}\gamma_{\rho})=\frac{i}{2}\gamma_{\rho\sigma}.
\end{equation}

In this paper, we discuss the quantum microcausality of spin-noncommutative
quantum field theories. The microcausality of noncommutative quantum
field theories based on eq.(\ref{eq:canonc}) has been discussed by
several authors. In\cite{Chaichian:2002vw,Ma:2006qx}, the authors
have studied microcausality by computing the commutator of operators
of observables $\mathcal{O}=:\phi*\phi:$ at spacelike seperation
and found that microcausality is obeyed provided that $\theta^{0i}=0$.
But it is showed that the commutators for observables with partial
derivatives of fields do not vanish at spacelike seperation\cite{Greenberg:2005jq}.
Similar results are obtained by other authors \cite{Haque:2007rb}
in a different approach in noncommutative scalar and Yukawa theories.
The microcausility of noncommutative scalar field theory is revisited
in a more mathematical method-distributions theory in\cite{Soloviev:2008yb}
and the author proved that microcausality is violated for observables
$\mathcal{O}$ even for space-space noncommutativity. All the above
negative results are obtained for noncommutative field theories based
on eq.(\ref{eq:canonc}). Maybe we can see these results from a simple
viewpoint-they violate common Lorentz covariance. How about a Lorentz
covariant noncommutative field theory? Spin-noncommutative field theories
are suitable models for consideration. In section 2, we study two
examples in spin-noncommutative Dirac field theory.

\section{CALCULATION OF CAUSALITY}

For spin noncommutative algebras eq.(\ref{eq:spinnc1}) and eq.(\ref{eq:spinnc2})
of a massive Dirac spinor field with mass $m$, we have the following
representation \cite{Gomes:2010xk},

\begin{equation}
\hat{x}^{\mu}=x^{\mu}-\frac{i\theta}{2}\gamma^{5}\gamma^{\mu\nu}\partial_{\nu},\,\,\,\,\,\,\,\,\,\,\,\,\,\,\,\,\,\,\,\hat{p}^{\mu}=-i\partial^{\mu}.\label{eq:rep}
\end{equation}

Similar to Moyal product, one can define a new star product into which
we can encode the spin noncommutativiy 
\begin{equation}
(f*g)(x)=f(x)\exp(\frac{i\theta}{2}\overset{\leftharpoonup}{\partial}_{\mu}\gamma^{5}\gamma^{\mu\nu}\overset{\rightharpoonup}{\partial}_{\nu})g(x).\label{eq:star}
\end{equation}
In the expansion of Dirac fields, we use the normalization and convention
as 

\begin{eqnarray*}
\psi(x) & = & \int\frac{d^{3}p}{(2\pi)^{3/2}}\frac{1}{\sqrt{2\omega_{\mathrm{p}}}}\sum_{s}(\hat{b}_{\mathrm{p},s}u(\mathrm{p},s)e^{-ip\cdot x}+\hat{d}_{\mathrm{p},s}^{+}v(\mathrm{p},s)e^{ip\cdot x}),\\
\bar{\psi}(x) & = & \int\frac{d^{3}p}{(2\pi)^{3/2}}\frac{1}{\sqrt{2\omega_{\mathrm{p}}}}\sum_{s}(\hat{b}_{\mathrm{p},s}^{+}\bar{u}(\mathrm{p},s)e^{ip\cdot x}+\hat{d}_{\mathrm{p},s}\bar{v}(\mathrm{p},s)e^{-ip\cdot x}).
\end{eqnarray*}
where $\omega_{\mathrm{p}}=\sqrt{\mathrm{p}^{2}+m^{2}}$ and $u,v$
are the positive and negative energy solutions of Dirac equation.
The creation and annihilation operators obey anticommutation relations
\[
\{\hat{b}_{\mathrm{p},s},\hat{b}_{\mathrm{q},r}^{+}\}=\delta^{3}(\mathrm{p}-\mathrm{q})\delta_{sr},\,\,\,\,\,\,\,\,\,\,\,\{\hat{d}_{\mathrm{p},s},\hat{d}_{\mathrm{q},r}^{+}\}=\delta^{3}(\mathrm{p}-\mathrm{q})\delta_{sr}.
\]

To study microcausality, we choose Hermitian operator $\mathcal{O}(x)=:\bar{\psi}(x)*\psi(x):$
as a sample observable. First, we consider the vacuum expectation
value $\langle0|[\mathcal{O}(x),\mathcal{O}(y)]|0\rangle$. 

Using the definition of the star product, the observable $\mathcal{O}(x)$
can be written in explicit form
\begin{eqnarray}
\mathcal{O}(x) & = & \int\frac{d^{3}p_{1}}{(2\pi)^{3/2}}\frac{1}{\sqrt{2\omega_{\mathrm{p_{1}}}}}\int\frac{d^{3}p_{2}}{(2\pi)^{3/2}}\frac{1}{\sqrt{2\omega_{\mathrm{p_{2}}}}}\sum_{s_{1}}\sum_{s_{2}}\nonumber \\
 &  & (\hat{b}_{\mathrm{p_{1}}s_{1}}^{+}\hat{b}_{\mathrm{p}_{2}s_{2}}\bar{u}(\mathrm{p}_{1},s_{1})M(p_{1},p_{2})u(\mathrm{p}_{2},s_{2})e^{i(p_{1}-p_{2})\cdot x}\nonumber \\
 &  & +\hat{b}_{\mathrm{p_{1}}s_{1}}^{+}\hat{d}_{\mathrm{p}_{2}s_{2}}^{+}\bar{u}(\mathrm{p}_{1},s_{1})M(p_{1},-p_{2})v(\mathrm{p}_{2},s_{2})e^{i(p_{1}+p_{2})\cdot x}\nonumber \\
 &  & +\hat{d}_{\mathrm{p_{1}}s_{1}}\hat{b}_{\mathrm{p}_{2}s_{2}}\bar{v}(\mathrm{p}_{1},s_{1})M(-p_{1},p_{2})u(\mathrm{p}_{2},s_{2})e^{-i(p_{1}+p_{2})\cdot x}\nonumber \\
 &  & -\hat{d}_{\mathrm{p_{2}}s_{2}}^{+}\hat{d}_{\mathrm{p}_{1}s_{1}}\bar{v}(\mathrm{p}_{1},s_{1})M(-p_{1},-p_{2})v(\mathrm{p}_{2},s_{2})e^{-i(p_{1}-p_{2})\cdot x}),\label{eq:O(x)}
\end{eqnarray}
where the momentum-dependent matrix $M$ is defined as 
\begin{equation}
M(p,q)=\exp(\frac{i}{2}\theta\gamma^{5}\gamma^{\mu\nu}p_{\mu}q_{\nu}).\label{eq:M}
\end{equation}
From the definition, we can see that $M(p,q)=M(-p,-q),\,\,\,\,\, M(p,-q)=M(-p,q)$.
Then the vacuum expectation of the commutator of two observables are
\begin{eqnarray}
 &  & \langle0|[\mathcal{O}(x),\mathcal{O}(y)]|0\rangle=\int\frac{d^{3}p_{1}}{(2\pi)^{3/2}}\frac{1}{2\omega_{\mathrm{p_{1}}}}\int\frac{d^{3}p_{2}}{(2\pi)^{3/2}}\frac{1}{2\omega_{\mathrm{p_{2}}}}\nonumber \\
 &  & \lbrace\mathrm{tr}[(\slashed{p}{}_{1}+m)M(p_{1},-p_{2})(\slashed{p}{}_{2}-m)M(-p_{2},p_{1})](e^{-i(p_{1}+p_{2})(x-y)}-e^{i(p_{1}+p_{2})(x-y)})\rbrace.\label{eq:vv}
\end{eqnarray}
For the commutative limit, the matrix $M=1$, the above equation becomes
\begin{equation}
\int\frac{d^{3}p_{1}}{(2\pi)^{3/2}}\frac{1}{2\omega_{\mathrm{p_{1}}}}\int\frac{d^{3}p_{2}}{(2\pi)^{3/2}}\frac{1}{2\omega_{\mathrm{p_{2}}}}(4p_{1}\cdot p_{2}-4m^{2})(e^{-i(p_{1}+p_{2})(x-y)}-e^{i(p_{1}+p_{2})(x-y)}).\label{eq:comlim1}
\end{equation}
When $x-y$ is spacelike, we can take $x_{0}=y_{0}$ due to the Lorentz
invariance of the above equation. Then it is easy to see that eq.(\ref{eq:comlim1})
equals to zero. For noncommutative case, the difficulty in the integral
is the trace term in eq.(\ref{eq:vv}) , which can be reduced to 
\begin{equation}
\mathrm{tr}[\slashed{p}_{2}M(-p_{2},p_{1})\slashed{p}_{1}M(p_{1},-p_{2})-m^{2}].\label{eq:tr1}
\end{equation}
It is hard to get an explicit expression for the above trace, but
we know that the above trace is a Lorentz scalar. So it must be a
function of the form $F(p_{1}\cdot p_{2},m^{2})$ . Using the same
analysis of the commutative case, the eq.(\ref{eq:vv}) equals to
zero. We conclude that the microcausality is obeyed by vacuum expectation
of the commutator.

Then let us check the matrix element between vacuum and a two-particle
state. We choose the two particle state as $|q_{1},r_{1};q_{2},r_{2}\rangle=\hat{b}_{\mathrm{q}_{1}r_{1}}^{+}\hat{d}_{\mathrm{q}_{2}r_{2}}^{+}|0\rangle$
,
\begin{eqnarray}
 &  & \langle0|[\mathcal{O}(x),\mathcal{O}(y)]|q_{1},r_{1};q_{2},r_{2}\rangle=\frac{1}{2\sqrt{\omega_{q_{1}}\omega_{q_{2}}}}e^{-iq_{1}\cdot x-iq_{2}\cdot y}\int\frac{d^{3}p}{(2\pi)^{3/2}}\frac{1}{2\omega_{\mathrm{p}}}\lbrace(e^{-ip\cdot(x-y)}-e^{ip\cdot(x-y)})\nonumber \\
 &  & \bar{v}(q_{2},r_{2})[M(-q_{2},p)(\slashed{p}+m)M(p,q_{1})-M(-q_{2},-p)(\slashed{p}-m)M(-p,q_{1})]u(q_{1},r_{1})\rbrace.\label{eq:vt}
\end{eqnarray}
At this case, we choose a definite two-particle state, so the above
matrix element is not Lorentz invariant. For simplicity, the commutator
is taken to be equal time. It is conceivable that the above integral
is not zero and thus the microcausility is violated. But it is not
easy to obtain a explicitly analytic results from it. In order to
do this, we should take some limit and choose special spinor states
for the two particles. In chiral representation, the matrix in the
second line of the above equation
\[
[M(-q_{2},p)(\slashed{p}+m)M(p,q_{1})-M(-q_{2},-p)(\slashed{p}-m)M(-p,q_{1})]
\]
 can be written explicitly as

\[
\left(\begin{array}{cc}
m(e^{\vec{A}\cdot\vec{\sigma}}e^{\vec{C}\cdot\vec{\sigma}}+e^{-\vec{A}\cdot\vec{\sigma}}e^{-\vec{C}\cdot\vec{\sigma}}) & e^{\vec{A}\cdot\vec{\sigma}}(p_{0}-\vec{\mathrm{p}}\cdot\vec{\sigma})e^{\vec{D}\cdot\vec{\sigma}}-e^{-\vec{A}\cdot\vec{\sigma}}(p_{0}-\mathrm{p}\cdot\vec{\sigma})e^{-\vec{D}\cdot\vec{\sigma}}\\
e^{\vec{B}\cdot\vec{\sigma}}(p_{0}+\vec{\mathrm{p}}\cdot\vec{\sigma})e^{\vec{C}\cdot\vec{\sigma}}-e^{-\vec{B}\cdot\vec{\sigma}}(p_{0}+\vec{\mathrm{p}}\cdot\vec{\sigma})e^{-\vec{C}\cdot\vec{\sigma}} & m(e^{\vec{B}\cdot\vec{\sigma}}e^{\vec{D}\cdot\vec{\sigma}}+e^{-\vec{B}\cdot\vec{\sigma}}e^{-\vec{D}\cdot\vec{\sigma}})
\end{array}\right)
\]
where 
\begin{eqnarray}
A & = & \frac{i\theta}{2}(p^{0}\vec{\mathrm{q}}_{2}-q_{2}^{0}\vec{\mathrm{p}}+i\vec{\mathrm{q}}_{2}\times\vec{\mathrm{p}})\\
B & = & \frac{i\theta}{2}(p^{0}\vec{\mathrm{q}}_{2}-q_{2}^{0}\vec{\mathrm{p}}-i\vec{\mathrm{q}}_{2}\times\vec{\mathrm{p}})\\
C & = & \frac{i\theta}{2}(p^{0}\vec{\mathrm{q}}_{1}-q_{1}^{0}\vec{\mathrm{p}}+i\vec{\mathrm{q}}_{1}\times\vec{\mathrm{p}})\\
D & = & \frac{i\theta}{2}(p^{0}\vec{\mathrm{q}}_{1}-q_{1}^{0}\vec{\mathrm{p}}-i\vec{\mathrm{q}}_{1}\times\vec{\mathrm{p}}).
\end{eqnarray}
If we take the momenta of the particles to be zero, i.e. $q_{1}=q_{2}=(m,0,0,0)$
and the seperation to be $\vec{x}-\vec{y}=(0,0,z)$, it turns out
that the integral in eq.(\ref{eq:vt}) is zero. Microcausality is
preserved by this special case. Then we should consider other cases.
For the sake of simplicity, we take massless limit and the momenta
are chosen to be $q_{1}=q_{2}=(q,0,0,q)$ and the seperation to be
$\vec{x}-\vec{y}=(0,0,z)$. The spinor states of the two particles
are $u=v\propto\left(\begin{array}{c}
\eta\\
0
\end{array}\right),\eta=\left(\begin{array}{c}
1\\
0
\end{array}\right)$. Then the integral (up to an irrelevant coefficient) in eq.(\ref{eq:vt})
is
\begin{eqnarray}
\frac{1}{4(z^{2}-\theta^{2}q^{2})^{2}}\int_{0}^{\infty}\mathrm{dp}[-2i\theta qz\cos(z\mathrm{p})(e^{2i\theta q\mathrm{p}}+e^{-2i\theta q\mathrm{p}})+4i\theta qz\cos(z\mathrm{p})\nonumber \\
-(\theta^{2}q^{2}+z^{2})\sin(z\mathrm{p})(e^{2i\theta q\mathrm{p}}-e^{-2i\theta q\mathrm{p}})+4i\theta q(z^{2}-\theta^{2}q^{2})\mathrm{p}\sin(z\mathrm{p})]\nonumber \\
=\frac{\pi}{4(z^{2}-\theta^{2}q^{2})^{2}}[i(z-\theta q)^{2}\delta(z+2\theta q)+i(z+\theta q)^{2}\delta(z-2\theta q)+4i\theta qz\delta(z)\nonumber \\
-4i\theta q(z^{2}-\theta^{2}q^{2})\delta^{'}(z)].\label{eq:integral}
\end{eqnarray}
This matrix element is not zero. So microcausality is violated by
this case. This result is similar to the one in canonical noncommutative
models where the violation is also proportional to sum of $\delta$-functions
\cite{Greenberg:2005jq}. We can also see that for commutative limit,
$\theta\rightarrow0$, the integral eq.(\ref{eq:integral}) is zero
and microcausality is preserved as is expected. If $z\rightarrow0$,
the integral is zero too as it should be for a commutator of the same
operator.

\section{conclusion and discussion}

In this report, we discuss the problem of microcausality of an interesting
Lorentz invariant noncommutative field theory where the noncommutativity
is induced by the spin of the field. We take $\mathcal{O}(x)=:\bar{\psi}(x)*\psi(x):$
as a sample observable and show explicitly by two examples that microcausality
is violated for the theory in general. In the canonical noncommutative
theories with constant noncommutative parameters $\theta^{\mu\nu}$,
the noncommutativity appears as extra phase factors depending on external
or internal momenta. If there is only space-space noncommutativity\cite{Gomis:2000zz},
the microcausality is preserved by observables without time derivatives\cite{Greenberg:2005jq},
and for general observables, microcausality is violated. In this spin-induced
noncommutative model, the noncommutativity also depends on the momenta
of external or internal momenta. From eq.(\ref{eq:integral}), one
can see that the violation of the microcausality consists of sum of
$\delta$-functions and nonlocality is proportional to the momenta
of external particles. These features are similar to the ones found
in canonical Moyal-type noncommutative theories \cite{Greenberg:2005jq}.
In Moyal-type noncommutative theories, there are UV/IR mixing problems.
One interesting thing is to explore whether there are similar problems
in quantum field theories on this spin-induced noncommutative spacetime.

In classical special relativity and common quantum field theories,
Lorentz covariance preserves causality and guarantees that one can
not travel back through time. But on noncommutative spacetime, even
Lorentz covariance can not preserve causality. It is also noted that
another Lorentz-covariant noncommutative model is proposed by\cite{Carlson:2002wj}
and the Lorentz violation appears when noncommutative parameters integrated
over all possible values and directions\cite{Saxell:2008zj}. We emphasize
that this nonlocality spreading procedure don't appear in spin-induced
noncommutative theory. In both cases, Lorentz covariance of the action
doesn't mean microcausality or local commutativity. This fact supports
a viewpoint, proposed by Greenberg\cite{Greenberg:2004vt}, that the
Lorentz covariance of time-ordered product of fields leads to microcausality.
Anyway, it is a really nontrival work to construct a quantum theory
of spacetime to be consistent with causality. 
\begin{acknowledgments}
The authors thank V. Kupriyanov and A.J. da Silva for the communication
with them and help from them. This work is supported by the Fundamental
Research Funds for the Central Universities with the contract number
2009QNA3015.\end{acknowledgments}


\begin{thebibliography}{10}
\bibitem{snyder1947}H. Snyder, 
Phys. Rev. {\bf 71}, 38 (1947);
H. Snyder, 
Phys. Rev. {\bf 72}, 68 (1947).

\bibitem{Douglas2001}M. R. Douglas and N. A. Nekrasov, 
Rev. Mod. Phys. {\bf 73},977 (2001).

\bibitem[3]{Szabo2003}R. J. Szabo, 
Phys. Rep. {\bf 378}, 207 (2003).

\bibitem[4]{dou}M.R. Douglas and C. Hull, 
J. High Energy Phys. {\bf 02},(1998)008.

\bibitem[5]{cheung1998}Yeuk-Kwan E. Cheung and Morten Krogh , 
Nucl.Phys. B {\bf 528}, 185(1998).

\bibitem[6]{chu1999}C. S. Chu and P. M. Ho, 
Nucl.Phys.B {\bf 550}, 151(1999).

\bibitem[7]{seiberg1999}N. Seiberg and E. Witten, 
J. High Energy Phys. {\bf 09},(1999)032.

\bibitem[8]{moyal}J. E. Moyal, 
Proc.Cambridge Phil. Soc. {\bf 45}, 99(1949).

\bibitem[9]{Chaichian:2004yh}M. Chaichian, P. Kulish, K. Nishijima, and A. Tureanu, 
Phys. Lett. B {\bf 604},98(2004);
M.~Chaichian, P.~Presnajder and A.~Tureanu,   
Phys.\ Rev.\ Lett.\  {\bf 94}, 151602 (2005).


\bibitem[10]{Wess:2003da}J.~Wess,   
hep-th/0408080. 

\bibitem[11]{Falomir:2009cq}H.~Falomir, J.~Gamboa, J.~Lopez-Sarrion, F.~Mendez, P.~A.~G.~Pisani,   Phys.\ Lett.\  {\bf B680}, 384(2009).

\bibitem[12]{Gomes:2010xk}M.~Gomes, V.~G.~Kupriyanov and A.~J.~da Silva,   
Phys.\ Rev.\  D {\bf 81}, 085024 (2010). 

\bibitem[13]{Das:2011tj}A.~Das, H.~Falomir, M.~Nieto, J.~Gamboa, F.~Mendez,
Phys.\ Rev.\  {\bf D84}, 045002 (2011).

\bibitem[14]{Chaichian:2002vw}M.~Chaichian, K.~Nishijima and A.~Tureanu,
Phys.\ Lett.\  B {\bf 568}, 146 (2003). 

\bibitem[15]{Ma:2006qx}Z.~Z.~Ma,   
hep-th/0603054. 

\bibitem[16]{Greenberg:2005jq}O.~W.~Greenberg,   
Phys.\ Rev.\  D {\bf 73}, 045014 (2006). 

\bibitem[17]{Haque:2007rb}A.~Haque and S.~D.~Joglekar,   
J.\ Phys.\ A  {\bf 41}, 215402 (2008). 

\bibitem[18]{Soloviev:2008yb}M.~A.~Soloviev,   
Phys.\ Rev.\  D {\bf 77}, 125013 (2008);  
M.~A.~Soloviev,   
J.\ Phys.\ A  {\bf 40}, 14593 (2007). 

\bibitem[19]{Gomis:2000zz}J.~Gomis, T.~Mehen,   
Nucl.\ Phys.\  {\bf B591}, 265(2000).

\bibitem[20]{Carlson:2002wj}C.~E.~Carlson, C.~D.~Carone, N.~Zobin,   
Phys.\ Rev.\  {\bf D66}, 075001 (2002).   

\bibitem[21]{Saxell:2008zj}S.~Saxell,   
Phys.\ Lett.\  {\bf B666}, 486(2008). 

\bibitem[22]{Greenberg:2004vt}O.~W.~Greenberg,   
Phys.\ Rev.\  {\bf D73}, 087701 (2006).  \end{thebibliography}
\end{document}